\documentclass[notitlepage, 12pt]{article}

\usepackage{color}
\usepackage{latexsym}
\usepackage{amsmath, amssymb, amsfonts}
\usepackage[toc]{appendix}
\usepackage[makeroom]{cancel}
\numberwithin{equation}{section}

\newtheorem{lemma}{Lemma}

\newtheorem{theorem}{Theorem}
\newtheorem{corollary}{Corollary}

\newtheorem{proposition}{Proposition}

\newtheorem{remark}{Remark}

\newcommand{\beginsec}{
\setcounter{lemma}{0}
\setcounter{theorem}{0}
\setcounter{corollary}{0}
\setcounter{definition}{0}
\setcounter{example}{0}
\setcounter{proposition}{0}
\setcounter{condition}{0}
\setcounter{assumption}{0}
\setcounter{conjecture}{0}
\setcounter{problem}{0}
\setcounter{remark}{0}
}

\newcommand{\noi}{\noindent}

\newcommand{\E}{\mathbb{E}}
\newcommand{\R}{\mathbb{R}}

\newcommand{\la}{\lambda}
\newcommand{\al}{\alpha}
\newcommand{\kap}{\kappa}
\newcommand{\sig}{\sigma}

\newcommand{\gam}{\gamma}
\newcommand{\del}{\delta}

\newcommand{\FF}{{\mathbb F}}
\newcommand{\PP}{{\mathbb P}}

\newcommand{\calC}{{\cal C}}

\newcommand{\calF}{{\cal F}}

\newcommand{\calL}{{\cal L}}

\newcommand{\skp}{\vspace{\baselineskip}}

\newcommand{\qed}{\hfill $\Box$ \skp}

\newcommand {\pf} {\noi \textit{Proof.} }

\title{Minimizing the Expected Lifetime Spent in Drawdown under Proportional Consumption}

\author{Bahman Angoshtari\thanks{email: bango@umich.edu} \hspace{4em} Erhan Bayraktar\thanks{email: erhan@umich.edu, E.\ Bayraktar thanks the National Science Foundation for financial support under grant number DMS-0955463.} \hspace{4em} Virginia R.\ Young\thanks{email: vryoung@umich.edu, V.\ R.\ Young thanks the Cecil J.\ and Ethel M.\ Nesbitt Professorship for financial support.}\\ \\ \\
Department of Mathematics\\
University of Michigan\\
Ann Arbor, 48109, USA}

\date{August 21, 2015}

\begin{document}

\maketitle

\begin{abstract}
We determine the optimal amount to invest in a Black-Scholes financial market for an individual who consumes at a rate equal to a constant proportion of her wealth and who wishes to minimize the expected time that her wealth spends in drawdown during her lifetime.  Drawdown occurs when wealth is less than some fixed proportion of maximum wealth.  We compare the optimal investment strategy with those for three related goal-seeking problems and learn that the individual is myopic in her investing behavior, as expected from other goal-seeking research.
\end{abstract}

\noi {\bf JEL subject classifications.} C61, G02, G11.  \smallskip

\noi {\bf AMS subject classifications.} Primary, 49L20, 60H30; Secondary, 35Q93  \smallskip

\noi {\bf Key words.}  Drawdown, occupation time, optimal investment, stochastic control, free-boundary problem.

\section{Introduction}\label{sec1}
\beginsec

Drawdown occurs when the value of an investor's portfolio drops to a fixed proportion of its maximum value.  Angoshtari et al.\ \cite{ABY2015a} and Chen et al.\ \cite{CLLL2015} computed the optimal investment strategy to minimize the probability that drawdown occurs during the investor's life.  The problem they considered essentially ends if drawdown occurs; however, when drawdown occurs, the individual must continue investing and consuming.  Thus, in this paper, we determine the investment strategy to minimize the expected {\it time} that the individual's wealth spends in drawdown during her life.  Bayraktar and Young \cite{BY2010} solved the problem of minimizing the expected time that an individual's wealth spends in a specific interval, namely, $[-L, 0]$ for $L > 0$ large, during her life; that is, they minimized expected lifetime {\it occupation}.  The work in this paper differs from \cite{BY2010} in that the individual controls the interval of occupation (specifically, the region of drawdown) by controlling maximum wealth.

In most research involving drawdown, wealth is constrained not to experience drawdown; see Grossman and Zhou \cite{GZ1993} and Cvitani\'c and Karatzas \cite{CK1995} for early references, and see Kardaras et al.\ \cite{KOP2014} for a recent reference.  However, if the individual is consuming continually from her investment account, then one cannot prevent drawdown, so minimizing the expected time spent in drawdown is a reasonable, objective goal.

In a related paper, Zhang \cite{Zhang2015} considered the drawdown of a one-dimensional, time-homogeneous diffusion, in which he defined drawdown as wealth dropping below its maximum by a {\it constant} amount, as opposed to what one might call {\it relative} drawdown in this paper.  In Section 4.5 of \cite{Zhang2015}, the author computed the Laplace transform of the time that the diffusion spent in drawdown until an independent exponential time.  However, Zhang \cite{Zhang2015} did not control the diffusion, as we do.

The remainder of this paper is organized as follows.  In Section \ref{sec2}, we describe the financial model and define (life)time spent in drawdown.  In Section 3, we compute the minimum expectation of lifetime spent in drawdown and compare the optimal investment strategy with those for three related goal-seeking problems.  In our main result, Theorem \ref{thm:pi}, we show that the individual is myopic in her investing behavior, as expected from other goal-seeking research.

\section{The model}\label{sec2}
\beginsec

In Section \ref{subsec21}, we describe the financial market in which the individual invests, and we define the value function that measures expected lifetime spent in drawdown.  Then, in Section \ref{sec2a}, we present a verification lemma that we use in Section \ref{sec4} to solve the investor's control problem.

\subsection{Background and statement of problem}\label{subsec21}

We assume the investor trades continuously in a Black--Scholes market with no transaction costs.  Borrowing and short selling are allowed.  The market consists of two assets, a riskless asset and a risky asset. The riskless asset earns interest at the constant rate $r > 0$.  The price of the risky asset follows geometric Brownian motion given by
\begin{equation*}
dS_t = S_t\left(\mu dt + \sig dB_t\right),
\end{equation*}
in which $\mu > r$, $\sig > 0$, and $(B_t)_{t\ge 0}$ is a standard Brownian motion on a filtered probability space $(\Omega, \calF, \FF = \{ \calF_t \}_{t \ge 0}, \PP)$, in which $\mathcal{F}_t$ is the augmentation of $\sig(B_u\;:\; 0\le u\le t)$.\footnote{If the drift of the risky asset $\mu$ is less than the riskless rate $r$, then the individual will optimally invest a negative amount in the risky asset, that is, she will ``short'' it.  Thus, to keep investment in the risky asset positive, we assume that $\mu > r$.  Also, investors want to receive greater return as they take on more risk, so we assume $\mu > r$ because $\sig > 0$.}

Let $W_t$ denote the wealth of the individual's investment account at time $t \ge 0$, and let $\pi_t$ denote the dollar amount invested in the risky asset at time $t \ge 0$.  An investment policy $\{ \pi_t \}_{t \ge 0}$ is {\it admissible} if it is an $\FF$-progressively measurable process satisfying $\int_0^t \pi^2_s \, ds < \infty$ almost surely, for all $t \ge 0$.

We assume that the investor's (net) consumption rate is proportional to her wealth, that is, it equals $\kap w$ when wealth equals $w$, with $\kap > r$.  Then, the wealth process follows the dynamics
\begin{equation*}
               dW_t = \left[ - (\kap - r)W_t + (\mu-r)\pi_t   \right]dt + \sig \pi_t dB_t, \quad t \ge 0,
\end{equation*}
in which $W_0 = w > 0$.  If wealth reaches $0$, we treat $0$ as an absorbing state for wealth, so that $W_t = 0$ for all $t \ge \inf \{ s: W_s \le 0 \}$.

Define maximum wealth $M_t$ at time $t$ by
\begin{equation*}
M_t = \max \left[ \sup_{0 \le s \le t} W_s, \; M_0 \right],
\end{equation*}
\noindent in which we include $M_0 = m \ge w$ (possibly different from $W_0 = w$) to allow the individual to have a financial past.  By {\it time spent in drawdown}, we mean the time the individual's wealth spends between $0$ and $\al M_t$, in which $\al \in (0, 1)$.  Specifically, denote by $X_t$ the time spent in drawdown on or before time $t$, so
\begin{equation*}
X_t = X_0 + \int_0^t {\bf 1}_{\{ W_s \le \al M_s \}} \, ds,
\end{equation*}
in which $X_0 = x \ge 0$.

By expected {\it lifetime spent in drawdown}, we mean the expected time the individual's wealth spends in drawdown before she dies. Specifically, we mean the expectation of $X_{\tau_d}$, in which $\tau_d$ is the random time of death of the investor.  We assume that $\tau_d$ is exponentially distributed with hazard rate $\la > 0$, that is, $\PP(\tau_d > t) = e^{- \la t}$.  If wealth reaches $0$, then the individual spends the remainder of her life in drawdown, with expected time $\frac{1}{\la}$.

Denote the minimum expected lifetime spent in drawdown by $\psi(w, m, x)$, in which the arguments $w$, $m$, and $x$ indicate that one conditions on the individual possessing wealth $w$ at the current time, with maximum (past) wealth $m$ and (previous) time spent in drawdown $x$.  Thus,
\begin{equation}
\label{eq:psi}
\psi(w, m, x) = \inf_{\{\pi_t \}} \E^{w, m, x}  \left(X_{\tau_d} \right),
\end{equation}
in which we minimize over admissible investment strategies, and $\E^{w, m, x}$ indicates that we condition the expectation on $W_0 = w$, $M_0 = m$, and $X_0 = x$.

\subsection{Verification lemma}\label{sec2a}

In this section, we provide a verification lemma that characterizes the value function $\psi$ as a unique solution to a boundary-value problem. We do not prove the theorem because its proof is similar to others in the literature; see, for example, \cite{WY2012}, \cite{CLLL2015}, or \cite{ABY2015a}.  Let
\begin{equation*}
D = \{ (w, m) \in (\R^+)^2: 0 \le w \le m \},
\end{equation*}
and for every $\pi \in \R$, define the following differential operator $\calL^\pi$ by
\begin{equation*}
\calL^\pi f = (-(\kap - r)w + (\mu-r) \pi) f_w + \frac{1}{2}\sig^2 \pi^2 f_{ww} - \la f + {\bf 1}_{\{ w \le \al m \}},
\end{equation*}
in which $f$ is a twice-differentiable function with respect to $w$.
 
\begin{lemma}\label{lem:verf}
Let $\phi = \phi(w, m)$ be a $\calC^{2, 1}$ function on $D$ $($except perhaps at $w = \al m$, where it will be $\calC^{1, 1}$ and have left and right second $w$-derivatives$)$ that is decreasing and convex with $w$ and increasing with $m$.  Suppose $\phi$ solves the following boundary-value problem.\footnote{By $\phi \in \calC^{2, 1}$, we mean that $\phi$ is $\calC^2$ with respect to its first argument $w$ and $\calC^1$ with respect to its second argument $m$; similarly, for $\calC^{1, 1}$.}
\begin{align}
\label{eq:BVP}
\left\{\begin{array}{ll}
               \displaystyle \inf_{\pi} \calL^\pi \phi(w, m) = 0, \quad 0 < w < m,
\\ \\
              \phi(0, m) = \dfrac{1}{\la}, \quad \displaystyle \lim_{w \to m-} \phi_{ww}(w, m) = +\infty.
              \end{array}
\right.
\end{align}
Then, the minimum expected lifetime spent in drawdown $\psi$ in \eqref{eq:psi} equals $\psi(w, m, x) = \phi(w, m) + x$ on $D \times \R^+$, and the optimal amount invested in the risky asset is given in a feedback form by
\begin{equation}
\label{eq:pi}
\pi^*_t = -\frac{\mu-r}{\sig^2} \, \frac{\phi_w(W^*_t, M^*_t)}{\phi_{ww}(W^*_t, M^*_t)},
\end{equation}
for $t \in [0, \tau_d)$, in which $W^*$ and $M^*$ are optimally controlled wealth and maximum wealth, respectively.
\end{lemma}

\begin{remark}
The condition $\displaystyle \lim_{w \to m-} \phi_{ww}(w, m) = +\infty$ implies that the amount invested in the risky asset approaches $0$ as wealth approaches the current maximum, which prevents maximum wealth from increasing because of the resulting negative drift and zero volatility in the wealth process as $w \to m-$.  For our problem, as well as for the related one considered by Chen et al.\ \cite{CLLL2015} of minimizing the {\rm probability} of lifetime drawdown under consumption proportional to wealth, it is optimal for maximum wealth {\rm not} to increase above its current maximum.  That is, $M^*_t = m$ with probability $1$, for all $t \ge 0$.  Intuitively, if maximum wealth were to increase, then the drawdown level would increase, too, so that spending time in drawdown would become more likely.  Thus, it is optimal not to allow maximum wealth to increase. 

This behavior may only happen for proportional consumption; for example, in minimizing the probability of lifetime drawdown under constant consumption $c > 0$, if maximum wealth is close enough to the so-called {\rm safe level} $c/r$, then it is optimal to allow maximum wealth to increase to the safe level; see Angoshtari et al.\ \cite{ABY2015a}.  For an individual minimizing expected lifetime spent in drawdown under constant consumption, we also expect her to allow maximum wealth to increase if it is close enough to the safe level.  On the other hand, in the set up in this paper, there is no safe level because $\kap > r$.
\end{remark}

Lemma \ref{lem:verf} allows us to reduce our three-dimensional problem to a two-dimensional one.  Because consumption is proportional to wealth, we can further reduce the dimension of the problem, as stated in the following corollary.  

\begin{corollary}
\label{cor:verf}
Let $\zeta = \zeta(z)$ be a $\calC^2$ function on $[0, 1]$ $($except perhaps at $\al$, where it will be $\calC^1$ and have left and right second derivatives$)$ that is decreasing and convex.  Suppose $\zeta$ solves the following boundary-value problem
\begin{align}
\label{eq:BVPzeta}
\left\{\begin{array}{ll}
               \displaystyle \la \zeta = -(\kap - r)z \, \zeta_z - \del \, \frac{\zeta_z^2}{\zeta_{zz}} + {\bf 1}_{\{ z \le \al \}}, \quad 0 < z < 1,
\\ \\
              \zeta(0) = \dfrac{1}{\la}, \quad \displaystyle \lim_{z \to 1-} \zeta_{zz}(z) = +\infty,
              \end{array}
\right.
\end{align}
in which
\begin{equation}
\label{eq:del}
\del = \frac{1}{2} \left( \frac{\mu - r}{\sig} \right)^2.
\end{equation}
Then, the minimum expected lifetime spent in drawdown $\psi$ in \eqref{eq:psi} is given by
\begin{equation}
\label{eq:psiz}
\psi(w, m, x) = \zeta(w/m) + x,
\end{equation}
on $D \times \R^+$, and the optimal amount invested in the risky asset when $W^*_t = w$ and $M^*_t = m$ is given by
\begin{equation}
\label{eq:pizeta}
\pi^*(w, m) = -\frac{\mu-r}{\sig^2} \; \frac{m \, \zeta_z(w/m)}{\zeta_{zz}(w/m)},
\end{equation}
independent of the time spent in drawdown $X^*_t = x$.
\end{corollary}

\section{Minimum expected lifetime spent in drawdown}\label{sec4}

In this section, we solve the optimization problem in \eqref{eq:psi}.  In Section \ref{sec4a}, we first analyze an auxiliary free-boundary problem (FBP) and then, in Section \ref{sec4b}, connect the solution of the FBP to the expected lifetime spent in drawdown via the Legendre transform.  In Section \ref{sec4c}, we study properties of the optimal investment strategy; in particular, we compare it with the optimal investment strategies for three related goal-seeking problems.

\subsection{Related free-boundary problem}\label{sec4a}

Consider the following FBP on $[y_1, \infty)$, with $0 < y_1 < y_\al < \infty$ to be determined.  \smallskip
\begin{align}
\label{eq:FBP}
\left\{\begin{array}{ll}
               \displaystyle \la \hat{\zeta}(y) = -(r - \kap - \la) y \hat{\zeta}(y) +  \del y^2\hat{\zeta}_{yy}(y) + {\bf 1}_{\{y \ge y_\al\}},
\\ \\
                \hat{\zeta}_y(y_1) = 1, \quad \hat{\zeta}_{yy}(y_1) = 0,
\\ \\
\hat{\zeta}_y(y_\al) = \al, \quad \displaystyle \lim_{y \to +\infty} \hat{\zeta}(y) = \frac{1}{\la}.
              \end{array}
\right.
\end{align}
In the following proposition, we present the solution of this FBP.

\begin{proposition}
\label{prop41}
The solution of the free-boundary problem \eqref{eq:FBP} on $[y_1, \infty)$ is given by
\begin{align}
\label{402}
\hat{\zeta}(y) = \left\{\begin{array}{ll}
               \displaystyle  \frac{y_1}{\gam_1 - \gam_2} \left[ \frac{1 - \gam_2}{\gam_1} \left( \frac{y}{y_1} \right)^{\gam_1} - \frac{1- \gam_1}{\gam_2}   \left( \frac{y}{y_1} \right)^{\gam_2} \right], \quad  &y_1 \le y < y_\al,
\\ \\
               \dfrac{1}{\la} + \dfrac{\al \, y_\al}{\gam_2}  \left( \dfrac{y}{y_\al} \right)^{\gam_2}, \quad &y \ge y_\al, 
              \end{array}
\right.
\end{align}
in which
\begin{align}
\label{eq:gam1}
\gam_1& = \frac{1}{2 \del} \left[(r - \kap - \la + \del) + \sqrt{(r - \kap - \la + \del)^2 + 4 \la \del} \, \right] \in (0,1),\\
\label{eq:gam2}
\gam_2& =  \frac{1}{2 \del} \left[(r - \kap - \la + \del) - \sqrt{(r - \kap - \la + \del)^2 + 4\la \del} \, \right] < 0.
\end{align}  \smallskip
The ratio of the free boundaries $y_{1 \al} := \dfrac{y_1}{y_\al} \in (0, 1)$ uniquely solves
\begin{equation}
\label{eq:y1al}
\frac{1 - \gam_2}{\gam_1 - \gam_2} \; y_{1 \al}^{1 - \gam_1} - \frac{1 - \gam_1}{\gam_1 - \gam_2} \; y_{1 \al}^{1 - \gam_2} = \al,
\end{equation}
the free boundary $y_{\al} > 0$ can be expressed in terms of $y_{1 \al}$ as
\begin{equation}
\label{eq:yal}
y_{\al} = \frac{1}{\la} \left( \frac{1 - \gam_2}{\gam_1(\gam_1 - \gam_2)} \,  y_{1 \al}^{1 - \gam_1} - \frac{1 - \gam_1}{\gam_2(\gam_1 - \gam_2)} \,  y_{1 \al}^{1 - \gam_1} - \frac{\al}{\gam_2} \right)^{-1},
\end{equation}
and the free boundary $y_1 = y_\al \cdot y_{1 \al}$.

Moreover, $\hat{\zeta}$ is increasing, concave, and $\mathcal{C}^2$, except at $y = y_\al$, where it is $\mathcal{C}^1$ and has left and right second derivatives.
\end{proposition}

\pf First, note that there exists a unique solution $y_{1 \al} \in (0, 1)$ of \eqref{eq:y1al}. Indeed, the left side of \eqref{eq:y1al} increases with respect to $y_{1 \al} \in (0, 1)$; when $y_{1 \al} = 0$, the left side of \eqref{eq:y1al} equals $0 < \al$; and, when $y_{1 \al} = 1$, the left side equals $1 > \al$.\footnote{Because of these properties, solving \eqref{eq:y1al} for $y_{1 \al}$ numerically is quite stable.}

Next, it is straightforward to show that the expression in \eqref{402} satisfies the differential equation in \eqref{eq:FBP} and that it satisfies the free-boundary conditions $\hat \zeta_y(y_1) = 1$ and $\hat \zeta_{yy}(y_1) = 0$, as well as the boundary condition $\displaystyle \lim_{y \to +\infty} \hat \zeta(y) = 1/\la$.  The expression in \eqref{eq:y1al} implies that $\hat \zeta$ in \eqref{402} satisfies the free-boundary condition $\hat \zeta_y(y_\al) = \al$; similarly, the expression in \eqref{eq:yal} implies that $\hat \zeta$ is continuous at $y = y_\al$. Thus, $\hat \zeta$ is $\calC^1$ at $y = y_\al$.  Also, note that $y_\al$ given in \eqref{eq:yal} is positive because all three terms in the parentheses are positive.

Finally, we show that $\hat \zeta$ given in \eqref{402} is increasing and concave. To that end, observe that
\begin{align*}
\hat{\zeta}_y(y) = \left\{
\begin{array}{ll}
               \displaystyle  \frac{1 - \gam_2}{\gam_1 - \gam_2} \left( \frac{y}{y_1} \right)^{\gam_1 - 1} - \frac{1- \gam_1}{\gam_1 - \gam_2}   \left( \frac{y}{y_1} \right)^{\gam_2 - 1}, \quad  &y_1 \le y < y_\al,
\\ \\
                \al \left( \dfrac{y}{y_\al} \right)^{\gam_2 - 1}, \quad &y \ge y_\al, 
\end{array}
\right.
\end{align*}
and
\begin{align}
\label{eq:hatzetayy}
\hat{\zeta}_{yy}(y) = \left\{
\begin{array}{ll}
               \displaystyle  - \, \frac{(1 - \gam_1)(1 - \gam_2)}{\gam_1 - \gam_2} \left[ \left( \frac{y}{y_1} \right)^{\gam_1 - 2} - \left( \frac{y}{y_1} \right)^{\gam_2 - 2} \right] \frac{1}{y_1}, \quad  &y_1 \le y < y_\al,
\\ \\
                - \, \dfrac{\al(1 - \gam_2)}{y_\al} \left( \dfrac{y}{y_\al} \right)^{\gam_2 - 2}, \quad &y > y_\al.
\end{array}
\right.
\end{align}
Because $\hat \zeta_{yy}(y) < 0$ for all $y \ge y_1$ ($y \ne y_\al$) and $\displaystyle \lim_{y \to +\infty} \hat \zeta_y(y) = 0$, we conclude that $\hat \zeta$ given in \eqref{402} is increasing and concave on $[y_1, \infty)$.  \qed

\subsection{Relation between the FBP and the expected lifetime spent in drawdown}\label{sec4b}

In this section, we prove that the Legendre transform of the solution \eqref{402} of the FBP \eqref{eq:FBP} satisfies the BVP \eqref{eq:BVPzeta} and, thereby, provides an implicit expression for the minimum expected lifetime spent in drawdown.  Because $\hat{\zeta}$ is concave, we can define its convex dual, as in the following proposition.  We abuse notation slightly by using $\zeta$ to denote the Legendre transform of $\hat \zeta$, but, as we will see, $\zeta$ thus defined satisfies the conditions of Corollary \ref{cor:verf} and solves \eqref{eq:BVPzeta}.

\begin{proposition}
\label{prop42}
Define the Legendre transform of $\hat \zeta$ by 
\begin{align}
\label{eq:Leg}
\zeta(z) := \max_{y \ge y_1} \left( \hat{\zeta}(y) - yz \right), \quad z \in [0, 1].
\end{align}
Then, $\zeta$ solves the boundary-value problem \eqref{eq:BVPzeta} and is decreasing, convex, and $\calC^2$, except at $z = \al$, where it is $\mathcal{C}^1$ and has left and right second derivatives.
\end{proposition}

\pf The optimizer $y = y(z)$ of the right side of \eqref{eq:Leg} solves the first-order condition $\hat \zeta_y(y) - z = 0$; thus, $y(z) = I(z)$, in which $I$ is the function inverse of $\hat \zeta_y$.  It follows that
\begin{equation}
\label{eq:zetaLeg}
\zeta(z) = \hat \zeta(I(z)) - z I(z).
\end{equation}
\eqref{eq:zetaLeg} implies that $\zeta_z(z) = I(z)$; thus, $y(z) = \zeta_z(z)$.  Moreover, $\zeta_z(z) = I(z)$ implies that $\zeta_{zz}(z) = -1/\hat \zeta_{yy}(I(z))$, from which it follows that $\zeta$ is decreasing and convex on $[0, 1]$.  Clearly, $\zeta$ is $\calC^2$, except at $z = \al$, where it is $\mathcal{C}^1$ and has left and right second derivatives.\footnote{$z = \al$ is ``dual'' to $y = y_{\al}$ because the slope of $\hat \zeta$ equals $\al$ when $y = y_{\al}$.}

By using these relationships and by substituting $y = I(z) = \zeta_z(z)$ into $\hat \zeta$'s FBP \eqref{eq:FBP}, we deduce that $\zeta$ solves the differential equation in \eqref{eq:BVPzeta}.  Next, $z = 0$ is ``dual'' to $y \to +\infty$ because the slope of $\hat \zeta$ approaches $0$ as $y$ approaches $+\infty$; thus, $\zeta(0) = \displaystyle \lim_{y \to +\infty} \hat \zeta(y) = 1/\la$.  Finally, $z = 1$ is ``dual'' to $y = y_1$ because the slope of $\hat \zeta$ equals $1$ when $y = y_1$; thus, $\displaystyle \lim_{z \to 1-} \zeta_{zz}(z) = - 1/\displaystyle \lim_{y \to y_1} \hat \zeta_{yy}(y) = + \infty$.  We have shown that $\zeta$ solves the BVP \eqref{eq:BVPzeta}.  \qed

The following theorem provides an implicit expression for the minimum expected lifetime spent in drawdown.
 
\begin{theorem}
\label{thm42}
The minimum expected lifetime spent in drawdown $\psi$ equals 
\begin{align}
\label{eq:psi_thm}
\psi(w, m, x) = x + \left\{\begin{array}{ll}
               \displaystyle \dfrac{1}{\la} + \dfrac{1 - \gam_2}{\gam_2} \, \al \, y_\al \, \left( \frac{w}{\al m} \right)^{- \frac{\gam_2}{1 - \gam_2}}, \quad &0 \le w \le \al m,
\\ \\
              y_1 \, \dfrac{(1 - \gam_1)(1 - \gam_2)}{\gam_1 - \gam_2} \left[ \dfrac{1}{\gam_1} \left( \dfrac{y}{y_1} \right)^{\gam_1} - \dfrac{1}{\gam_2} \left( \dfrac{y}{y_1} \right)^{\gam_2} \right], \quad &\al m < w \le m,
              \end{array}
\right.
\end{align}
and the optimal investment strategy is given in feedback form by $\pi^*_t = \pi^*(W^*_t, M^*_t)$, in which $W^*$ and $M^*$ are optimally controlled wealth and maximum wealth, respectively, and $\pi^*$ is defined by
\begin{align}
\label{eq:pi_thm}
\pi^*(w, m) = \left\{\begin{array}{ll}
               \displaystyle \frac{\mu-r}{\sig^2} \, (1 - \gam_2) \, w, \quad&  0 < w < \al m,
\\ \\
               \dfrac{\mu-r}{\sig^2} \, m \, \dfrac{(1 - \gam_1)(1 - \gam_2)}{\gam_1 - \gam_2} \left[ \left( \dfrac{y}{y_1} \right)^{\gam_1 - 1} - \left( \dfrac{y}{y_1} \right) ^{\gam_2 - 1} \right] ,\quad &\al m < w \le m.
              \end{array}
\right.
\end{align}
In the second expressions of \eqref{eq:psi_thm} and \eqref{eq:pi_thm}, $y \in [y_1, y_\al)$ uniquely solves 
\begin{equation}
\label{eq:yy1}
\frac{1 - \gam_2}{\gam_1 - \gam_2} \, \left( \frac{y}{y_1} \right) ^{\gam_1 - 1} -  \frac{1 - \gam_1}{\gam_1 - \gam_2} \,  \left( \frac{y}{y_1} \right) ^{\gam_2 - 1}  = \frac{w}{m},
\end{equation}
in which $y_1$ and $y_\al$ are defined in Proposition $\ref{prop41}$.
\end{theorem}

\pf We use Corollary \ref{cor:verf} to prove this theorem from Propositions \ref{prop41} and \ref{prop42}.  Indeed, because $\zeta$ defined in \eqref{eq:Leg} solves the BVP \eqref{eq:BVPzeta}, it follows from Corollary \ref{cor:verf} that $\psi(w, m, x) = \zeta(w/m) + x$ and $\pi^*(w, m) = - \frac{\mu - r}{\sig^2} \; m \, \zeta_z(w/m)/ \zeta_{zz}(w/m)$.

For $z \in [0, \al]$, we can explicitly compute the Legendre transform of $\hat \zeta$ to obtain
\begin{equation*}
\zeta(z) = \frac{1}{\la} + \frac{1 - \gam_2}{\gam_2} \, \al \, y_\al \left( \frac{z}{\al} \right)^{- \frac{\gam_2}{1 - \gam_2}};
\end{equation*}
thus, we have the first expressions for $\psi$ and $\pi^*$ in \eqref{eq:psi_thm} and \eqref{eq:pi_thm}, respectively.

For $z \in (\al, 1]$, we cannot explicitly compute the Legendre transform of $\hat \zeta$, but we have the implicit (second) expressions for $\psi$ and $\pi^*$ in \eqref{eq:psi_thm} and \eqref{eq:pi_thm}, respectively, in terms of the dual variable $y$.  Specifically, for $w \in (\al m, m]$, $y = y(z) \in [y_1, y_\al)$ uniquely solves $\hat \zeta_y(y) = w/m$, which is equivalent to \eqref{eq:yy1}. As a function of $y \in [y_1, y_\al)$, $\psi(w, m, x) = x + \hat{\zeta}(y) - y \cdot w/m$, which becomes the second expression in \eqref{eq:psi_thm} after we substitute for $w/m$ from \eqref{eq:yy1} and for $\hat \zeta(y)$ from the first expression in \eqref{402}.  Similarly, as a function of $y \in [y_1, y_\al)$, $\pi^*(w, m) = - \frac{\mu - r}{\sig^2} \; m \, y \, \hat \zeta_{yy}(y)$, which becomes the second expression in \eqref{eq:pi_thm} after we substitute for $\hat \zeta_{yy}(y)$ from the first expression in \eqref{eq:hatzetayy}.  \qed

\begin{remark}
Note that the optimal amount to invest in the risky asset is {\rm independent} of $y_1$.  Indeed, the solution $\frac{y}{y_1} \in [1, y_\al/y_1)$ of \eqref{eq:yy1} is independent of $y_1$, and once we have the ratio $\frac{y}{y_1}$, then the expression for $\pi^*(w, m)$ for $\al m < w \le m$ is independent of $y_1$.
\end{remark}

\subsection{The optimal investment strategy}\label{sec4c}

In this section, we compare $\pi^*$ with the optimal investment strategies for three related goal-seeking problems.  First, Young \cite{Young2004} showed that the optimal investment strategy to minimize the probability of lifetime ruin for a ruin level $w_r > 0$ under proportional consumption is given by
\begin{equation}
\label{eq:pir}
\pi^r(w) = \frac{\mu - r}{\sig^2} \, (1 - \gam_1) \, w,
\end{equation}
for $w \ge w_r$, in which $\gam_1$ is as given in \eqref{eq:gam1}.  Note that $\pi^r$ is independent of the ruin level $w_r$.

Second, Chen et al.\ \cite{CLLL2015} determined the optimal investment strategy to minimize the probability of lifetime drawdown under proportional consumption.  By using the method of Sections \ref{sec4a} and \ref{sec4b}, one can show that the optimal investment strategy for minimizing the probability of lifetime drawdown is given by
\begin{equation}
\label{eq:pid}
\pi^d(w, m) = \dfrac{\mu-r}{\sig^2} \, m \, \dfrac{(1 - \gam_1)(1 - \gam_2)}{\gam_1 - \gam_2} \left[ \left( \dfrac{y}{y_1} \right)^{\gam_1 - 1} - \left( \dfrac{y}{y_1} \right) ^{\gam_2 - 1} \right],
\end{equation}
for $w > \al m$, in which $\frac{y}{y_1} \ge 1$ solves \eqref{eq:yy1}.

Third, Cohen and Young \cite{CY2015} solved the problem minimizing the probability of ruin under poverty (with both constant and proportional consumption).  One can think of their problem as a generalization of minimizing the time that wealth occupies a given interval; see \cite{BY2010} for the solution of the occupation-time problem for constant consumption.  As a special case of \cite{CY2015}, the optimal investment strategy to minimize the expected occupation time of the interval $[0, \al m]$, with $m > 0$ fixed and independent of the wealth process, is given by
\begin{align}
\label{eq:pio}
\pi^o(w) = \left\{\begin{array}{ll}
               \displaystyle \frac{\mu-r}{\sig^2} \, (1 - \gam_2) \, w, \quad&  0 < w < \al m,
\\ \\
               \dfrac{\mu-r}{\sig^2} \, (1 - \gam_1) \, w ,\quad &w > \al m,
              \end{array}
\right.
\end{align}
in which $\gam_1$ and $\gam_2$ are given in \eqref{eq:gam1} and \eqref{eq:gam2}, respectively.

In the following theorem, our main result, we compare the optimal investment strategy $\pi^*$ given in \eqref{eq:pi_thm} with these three strategies.

\begin{theorem}
\label{thm:pi}
For $w_r \le w < \al m$ with $w_r > 0$ small,
\begin{equation}
\label{eq:comppi1}
\pi^*(w, m) = \pi^o(w) > \pi^r(w).
\end{equation}
For $\al m < w \le m$,
\begin{equation}
\label{eq:comppi2}
\pi^*(w, m) = \pi^d(w, m) < \pi^o(w) = \pi^r(w).
\end{equation}
In particular, $\pi^*(\al m-, m) > \pi^*(\al m+, m)$.
\end{theorem}

\pf  For $0 < w < \al m$, $\pi^*(w, m) > \pi^r(w)$, which follows from $1 - \gam_2 > 1 > 1 - \gam_1$.  For $\al m < w \le m$, $\pi^*(w, m) < \pi^r(w)$ is equivalent to
\begin{equation*}
\dfrac{(1 - \gam_1)(1 - \gam_2)}{\gam_1 - \gam_2} \left[ \left( \dfrac{y}{y_1} \right)^{\gam_1 - 1} - \left( \dfrac{y}{y_1} \right) ^{\gam_2 - 1} \right] < (1 - \gam_1) \, \frac{w}{m}.
\end{equation*}
After substituting for ${w}/{m}$ from \eqref{eq:yy1} and simplifying, this inequality is seen to be equivalent to $\gam_2 < \gam_1$, which is clearly true because $\gam_2 < 0 < \gam_1$.  The equalities in \eqref{eq:comppi1} and \eqref{eq:comppi2} are clear from the expressions for the four investment strategies.  \qed

When in drawdown, individuals who minimize expected time spent in drawdown or who minimize expected occupation time (of the same interval) invest identically, a type of myopic investment, which is not surprising because while an individual is in drawdown, nothing she does there will affect her maximum wealth, and the time penalty incurred while in drawdown or while in occupation is identical.  Both strategies ($\pi^*$ and $\pi^o$) are more heavily invested in the risky asset than when minimizing the probability of lifetime ruin because, while in drawdown, the former  continually incurs a time penalty, whereas the latter only incurs a penalty when ruin occurs.

When not in drawdown, individuals who minimize expected time spent in drawdown or who minimize the probability of drawdown occurring invest identically, another example of myopia.  This correspondence might seem, at first, surprising because the former does not incur a penalty until wealth spends a positive amount of time below $\al m$, but the latter incurs the maximum penalty the instant when wealth hits $\al m$.  However, if wealth hits $\al m$, then it will spend a positive amount of time below $\al m$ with probability $1$.

Also, when not in drawdown (or when not occupying $[0, \al m]$ in which we treat $\al m$ as independent of the occupier's wealth process),  individuals who minimize occupation time or who minimize the probability of ruin {\it for any positive ruin level} invest identically, a third example of myopia.  As in the previous paragraph, this correspondence might seem, at first, surprising because the former does not incur a penalty until wealth spends a positive amount of time below $\al m$, but the latter incurs the maximum penalty when wealth hits the ruin level (which we could set equal to $\al m$).  However, as observed in the previous paragraph, if wealth hits $\al m$, then it will spend a positive amount of time below $\al m$ with probability $1$.

Finally, when not in drawdown, $\pi^* = \pi^d$ are less than $\pi^o = \pi^r$ because the individual who minimizes expected time spent in drawdown invests so that maximum wealth does not increase above its current level, but the individual who minimizes occupation time in a fixed interval (independent of the wealth process) is happy if wealth increases to an arbitrarily large size.

We have observed myopic investment in other goal-seeking problems.  Bayraktar and Young \cite{BY2007} found the optimal investment strategy to minimize the probability of lifetime ruin under constant consumption and under a no-borrowing constraint on investment, that is, the individual was not allowed to invest more in the risky asset than her current wealth.  Under that constraint, when the constraint did {\it not} bind (specifically, at greater wealth levels), then the individual invested as if the constraint did not exist.  More recently, Bayraktar et al.\ \cite{BPY2015} and Bayraktar and Young \cite{BY2015} determined the optimal investment strategy to maximize the probability of reaching a bequest goal with and without life insurance, respectively.  In the wealth regions for which it is optimal {\it not} to buy life insurance (specifically, at lower wealth levels), then the individual invested as if life insurance were not available.  We conjecture that this myopia concerning constraints and opportunities is the rule, rather than the exception, in goal-seeking problems.

\bibliographystyle{plain}
\bibliography{bib_Jenny}

\end{document}